\documentclass[12pt,preprint]{aastex}
\usepackage{psfig,lscape}
\def\lap{\lower.5ex\hbox{$\; \buildrel < \over \sim \;$}}
\def\gap{\lower.5ex\hbox{$\; \buildrel > \over \sim \;$}}

\def\ergcm2s{${\rm erg\ cm^{-2}\ s^{-1}}$}

\def\ergscm2s{${\rm erg\ cm^{-2}\  s^{-1}}$}

\def\cm-2{${\rm cm^{-2}}$}
\def\ergs{${\rm erg\ s^{-1}}$}

\begin{document}

\title{Discovery of an X-ray Nova in M31}

\author{Benjamin F. Williams\altaffilmark{1}, Michael R. Garcia\altaffilmark{1}, Albert K. H. Kong\altaffilmark{1}, Frank A. Primini\altaffilmark{1}, and Stephen S. Murray\altaffilmark{1}}
\altaffiltext{1}{Harvard-Smithsonian Center for Astrophysics, 60
Garden Street, Cambridge, MA 02138; williams@head.cfa.harvard.edu;
garcia@head.cfa.harvard.edu;
akong@head.cfa.harvard.edu;
fap@head.cfa.harvard.edu;
rd@head.cfa.harvard.edu;
ssm@head.cfa.harvard.edu}


\begin{abstract}

We have obtained snapshot images of an X-ray nova in M31 from {\it
Chandra} ACIS-I and the {\it Hubble Space Telescope (HST)} Advanced
Camera for Surveys (ACS).  The {\it Chandra} position of the X-ray
nova was R.A.=00:44:06.68 $\pm$1.74$''$, Dec.=+41:12:20.0
$\pm$2.31$''$.  A follow-up {\it HST} observation 24 days later
revealed a source at R.A.=00:44:06.81, Dec.=+41:12:24.0 that was
$B=25.75\pm0.05$.  This optical source faded to $B=27.1\pm0.1$ in 3
months.  During this time period, the X-ray flux decayed linearly from
(3.6$\pm0.2)\times$10$^{-4}$ to $<$(6.9$\pm0.09)\times$10$^{-5}$ ct
cm$^{-2}$ s$^{-1}$. The {\it HST} identification of an optical source
in the same region experiencing an obvious drop in brightness in
concert with the X-ray nova suggests that this optical source is the
counterpart of the X-ray nova.  However, the precision of the X-ray
position allows the possibility that the optical source is a nearby
variable star.  We discuss the implications of both possibilities.

\end{abstract}

\section{Introduction}

Since the discovery of bright X-ray novae (XRNe), these bright,
transient X-ray events have allowed some of the most detailed studies
of the properties of the accretion disks of stellar-mass compact
objects (e.g. \citealp{chen1997,mcclintock2004}, and references
therein).  While XRNe have been observed in both high-mass X-ray
binaries (HMXBs; \citealp{tanaka1996}) and low-mass X-ray binaries
(LMXBs), most of those occurring in LMXBs are confirmed black hole
binaries.  In 15 Galactic XRNe systems optical followup has revealed
that the accreting object is more massive than $\sim 3$ M$_{\odot}$
\citep{mcclintock2004}, and therefore cannot be a stable neutron star
(NS).  The millisecond variability and $>10^{38}$\ergs\/ X-ray
luminosity further supports the argument that these binary systems
contain black holes.  Therefore, these objects are among the firmest
examples of black hole candidates known \citep{charles1998}, and are
of intense interest as sites to study general relativity in the
`strong gravity' regime.  The discovery of extragalactic XRNe is
therefore a crucial step forward in the study of extragalactic
stellar-mass black holes.

With X-ray luminosities ranging from 10$^{36}$--10$^{40}$ erg s$^{-1}$
\citep{chen1997}, XRNe can easily be detected by {\it Chandra} and
resolved from other X-ray sources within host galaxies several Mpc
away, making M31 a prime target for extragalactic studies of XRN with
{\it Chandra}.  In addition, with optical magnitudes ranging from
$-5\lap M_V \lap5$, it is possible to detect and resolve counterparts
for extragalactic XRNe in the Local Group with {\it HST}.

Extragalactic XRNe have also been discovered in the Magellanic Clouds
\citep{white1978}, M32 \citep{IAUC7498}, and M31
\citep{osborne2001,trudolyubov2001,atel76}.  {\it Chandra} and {\it
XMM} have allowed the discovery of nearly 50 XRNe in M31
\citep{trudolyubov2001,kong2002,distefano2004,williams2004}; Williams
et al. in preparation). Most recently, by combining the power of {\it
Chandra} with {\it HST}, X-ray/optical studies of the XRNe in M31 have
begun, providing new details and comparisons to their Galactic
siblings (e.g. \citealp{williams2004}).

Herein we report the discovery of a new XRN in M31 with new {\it
Chandra} ACIS-I data.  Along with three X-ray detections of this
source, we have obtained contemporaneous {\it HST} ACS images of the
region where the XRN appeared.  The X-ray data provide measurements of
the XRN luminosity and the decay pattern.  The optical data provide
constraints on the physical properties of the progenitor binary
system.

\section{Data}

\subsection{X-ray Data}

We obtained {\it Chandra} ACIS-I images of M31 on 09-Nov-2003,
26-Nov-2003, and 27-Dec-2003.  In addition to the 4 ACIS-I chips, the
S2 and S3 chips were on during the observations.  The observations
were performed in ``alternating exposure readout'', so that every 6th
frame had 0.6 seconds of exposure instead of the canonical 3.2
seconds.  This mode lowers the effective exposure time by $\sim$20\%,
but it provides a second low exposure image in which bright sources
are not piled up.  The details of these observations, including target
coordinates, roll angle of the telescope, and exposure time, are
provided in Table~\ref{xobs}.

These 3 observations were all reduced in an identical manner using the
X-ray data analysis package CIAO v3.0.2.  We created exposure maps for
the images using the CIAO script {\it
merge\_all},\footnote{http://cxc.harvard.edu/ciao/download/scripts/merge\_all.tar}
and we found and measured positions for the sources in the image using
the CIAO task {\it
wavdetect}.\footnote{http://cxc.harvard.edu/ciao3.0/download/doc/detect\_html\_manual/Manual.html}
Each data set detected sources down to (0.3--10 keV) fluxes of
$\sim$2$\times$10$^{-6}$ ct cm$^{-2}$ s$^{-1}$.  Errors for the X-ray
positions were determined using the IRAF\footnote{IRAF is distributed
by the National Optical Astronomy Observatory, which is operated by
the Association of Universities for Research in Astronomy, Inc., under
cooperative agreement with the National Science Foundation.} task {\it
imcentroid}, which determines the errors of the point spread function
(PSF) along each pixel axis independently.  The counts from the source
are projected onto each axis, and the error in the position is then
calculated by dividing the standard deviation
of the pixel positions of all of the source counts by the square root
of the number of counts.  As the pixels in {\it Chandra} images are
aligned with north up and east to the left, the X position error was
taken to be the R.A. error, and the Y position error was taken to be
the Dec. error.

We aligned the positions of bright ($>$100 counts) X-ray sources with
known globular cluster counterparts to the positions of the centers of
their host globular clusters in the images of the Local Group Survey
(LGS; \citealp{massey2001}) using the IRAF task {\it ccmap}.  These images have an assigned J2000
(FK5) world coordinate system accurate to $\sim$0.25$''$, and they
provided the standard coordinate system to which we aligned all of our
data for this project.  The alignment errors between the {\it Chandra}
and LGS are shown for each observation in column $\sigma_{AL}$ in
Table~\ref{xpos}.  Alignments allowed for adjustments in pixel scale
as well as shifts in $X$ and $Y$.

We cross-correlated the X-ray source positions of all 3 observations
against all previously published X-ray catalogs and the {\it
Simbad}\footnote{http://simbad.u-strasbg.fr/} database to look for any
new, bright X-ray source likely to be an XRN.

The data pertaining to the bright new X-ray source CXOM31
J004406.7+411220 (hereafter s1-86, see \citealp{williams2004} for
naming convention details) at R.A.= 00:44:06.68, Dec.=41:12:20.0
(J2000) was then studied in further detail.  Although the observations
were performed in the ACIS-I configuration, s1-86 was detected on the
S3 chip in the first 2 observations and on the S2 chip in the third
observation.

We extracted the X-ray spectrum of s1-86 from all 3 observations using
the CIAO task {\it
psextract}\footnote{http://cxc.harvard.edu/ciao/ahelp/psextract.html}.
We then fit these spectra independently, binning the spectra of the
first two observations so that each energy bin contained $\sim$20
counts and binning the spectrum of the third observation so that each
energy bin contained $\sim$10 counts.  These binning factors allowed
the use of standard $\chi^2$ statistics when fitting the spectra.

We fit the spectra with an absorbed power-law model and an absorbed
disk blackbody model using the CIAO 3.0/Sherpa fitting package
\citep{freeman2001}.  The best fitting model parameters, and the
associated fitting statistics, are provided in Tables~\ref{spec.dat}
and \ref{spec.dat2}.  Results are discussed in \S~\ref{results}.

\subsection{Optical Data}

We obtained two sets of {\it HST} ACS data, one observed at UT 21:35
on 03-Dec-2003 and one observed at UT 16:38 on 01-Mar-2004.  Each of
these were pointed at R.A.=00:44:07, Dec.=41:12:19.5.  The
observations had orientations of 71.75 deg and 32.73 deg respectively.
Both observations were taken using the standard ACS box 4-point dither
pattern to allow the final data to be drizzled to recover the highest
possible spatial resolution.  All exposures were taken through the
F435W filter.  The total exposure times were 2200 seconds for each
data set.

\begin{figure}
\centerline{\psfig{file=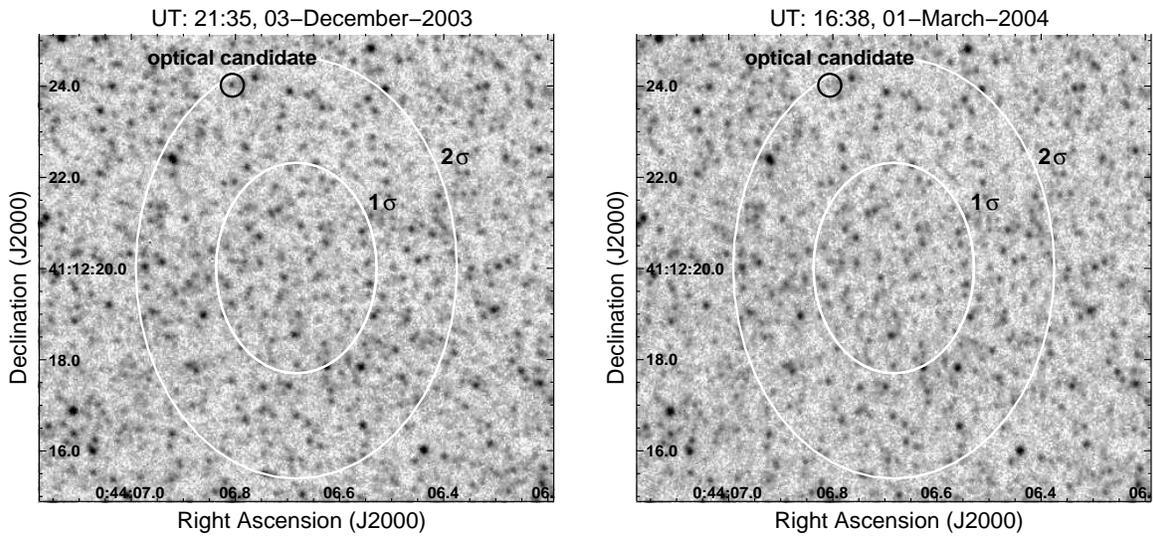,width=6.0in,angle=0}}
\caption{The relevant sections of the 2 drizzled ACS F435W images are
shown.  {\it Left panel:} the UT 21:35 03-Dec-2003 observation is
shown with a black circle marking the position of the source that
faded by more than 1 mag. {\it Right panel:} the UT 16:33 01-Mar-2004
observation is shown with a black circle marking the position of the
source that faded by more than 1 mag.  White ellipses mark the
1$\sigma$ ($\sim$8 pc) and 2$\sigma$ ($\sim$8 pc) X-ray position
errors.}
\label{hstims}
\end{figure}

We aligned and drizzled each set of 4 images into high-resolution
(0.025$''$ pixel$^{-1}$) images using the PyRAF\footnote{PyRAF is a
  product of the Space Telescope Science Institute, which is operated
  by AURA for NASA.} task {\it multidrizzle},\footnote{multidrizzle is
  a product of the Space Telescope Science Institute, which is
  operated by AURA for
  NASA. http://stsdas.stsci.edu/pydrizzle/multidrizzle} which has been
optimized to process ACS imaging data.  The task removes the cosmic
ray events and geometric distortions, and it drizzles the dithered
frames together into one final photometric image.  We processed the
relevant sections of the final images with DAOPHOT-II and ALLSTAR
\citep{stetson} to find and measure the count rates of the sources.
Sections (10$''$ $\times$ 10$''$) of the images, centered on the X-ray
position of the detected XRN, are shown in Figure~\ref{hstims}.

We converted the count rates measured on our images to ST magnitudes
and VEGA magnitudes using the conversion techniques provided in the
ACS Data
Handbook\footnote{http://www.stsci.edu/hst/acs/documents/handbooks/DataHandbookv2/ACS\_longdhbcover.html}.
The count rates, ST (F435W) magnitudes, and VEGA ($B$) magnitudes for
the optical source of interest are provided in Table~\ref{hst.dat} and
discussed in \S~\ref{pro}.

Finally, we aligned the drizzled ACS images with the LGS coordinate
system with {\it ccmap} using stars and clusters common to both
images.  This alignment was performed independently for each of the 2
ACS data sets to check the alignment consistency.  As displayed by the
coordinates in Figure~\ref{hstims} and the errors measured by {\it
ccmap} ($\sim$0.025$''$), we were able to accurately align the ACS
images to the LGS coordinate system.  Because the ACIS-I images had
been aligned to the same coordinate system, we were able to compare
the coordinates of objects the the ACS and ACIS-I images reliably.

\section{Results}\label{results}

\subsection{X-ray}

The first X-ray observation (ObsID 4678) provided the first detection
of s1-86 (CXOM31 J004406.7+411220 at R.A.= 00:44:06.71,
Dec.=41:12:20.0 (J2000)).  This position is 16$'$ east and 4$'$ south
of the nucleus, about 1$'$ east of the eastern spiral arm and in the
outskirts of the bulge.  Because this source was very bright and did
not appear in any previous publication, we assumed that it was an XRN.
We were able to find a 10 ks {\it XMM} image of this region
(OBS\_ID=0112570601) taken 28 December 2000, which contained this
region near the edge of the field.  Assuming a power-law spectrum with
a slope of 3.0 and an absorption column of 2$\times$10$^{21}$
cm$^{-2}$, the non-detection of s1-86 in this {\it XMM} observation
provides a 4$\sigma$ upper-limit to the unabsorbed 0.3--7 keV
luminosity of 2.5$\times$10$^{36}$ erg s$^{-1}$.  Our first detection
of the source, with an unabsorbed 0.3--7 keV luminosity of
2.4$\times$10$^{38}$ erg s$^{-1}$, shows that it brightened by a
factor of $\gap$100.

\begin{figure}
\centerline{\psfig{file=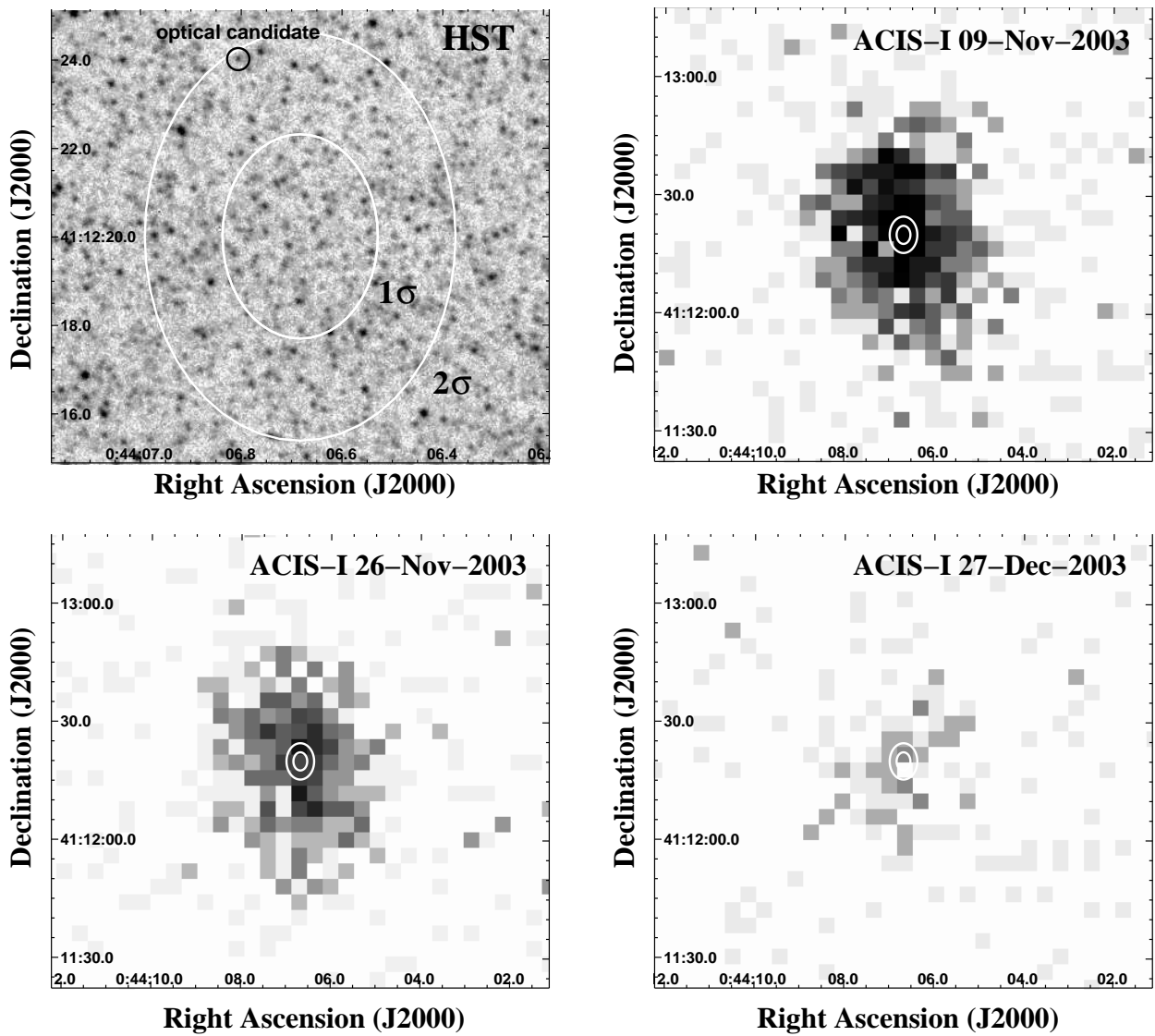,width=6.5in,angle=0}}
\caption{{\it Top left panel:} The 1$\sigma$ and 2$\sigma$ X-ray
position errors are shown with white ellipses on the {\it HST} image
from 03-Dec-2003.  The optical counterpart candidate is indicated with
the black circle.  {\it Top right panel:} The same error ellipses are
shown on the ACIS-I image of s1-86 from 09-Nov-2003.  {\it Lower left
panel:} The same error ellipses are shown on the ACIS-I image of s1-86
from 26-Nov-2003.  {\it Lower right panel:} The same error ellipses are
shown on the ACIS-I image of s1-86 from 27-Dec-2003.}
\label{xims}
\end{figure}

The next two observations in our M31 monitoring program (ObsIDs 4679
and 4680) confirmed the transient nature of the source, as the X-ray
flux decayed by a factor of 7.3 in 48 days.  This decay is clearly
seen in the X-ray images shown in Figure~\ref{xims}, demonstrating
that s1-86 is an XRN.

\begin{figure}
\centerline{\psfig{file=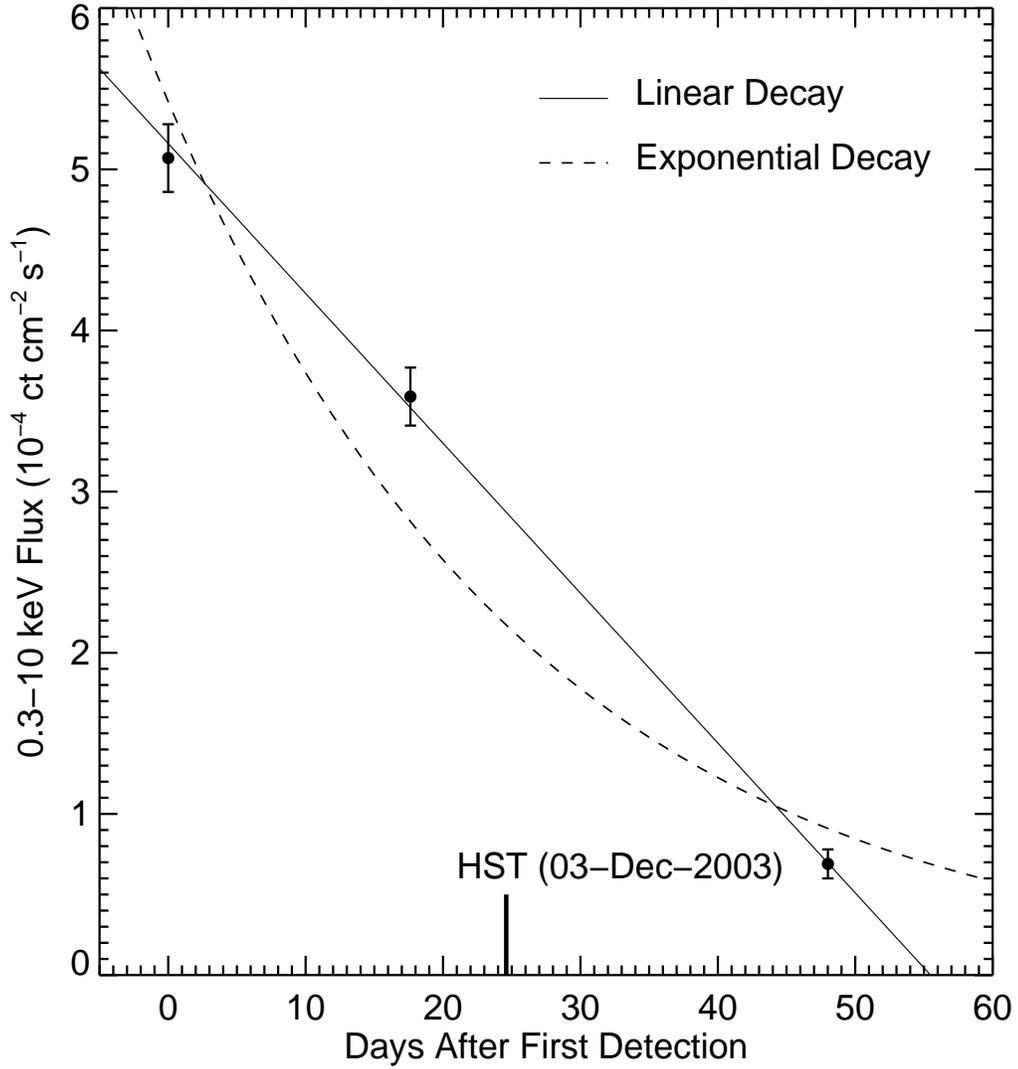,height=6.0in,angle=0}}
\caption{The lightcurve of s1-86 during our three ACIS-I detections.
{\it Dashed curve:} the best fit exponential decay curve, with an
$e$-folding time of 26.9 days ($\chi^2/\nu$ = 27.28/1).  {\it Solid
line:} the best fit linear decay curve, with a slope of $-$9.3
$\times$10$^{-6}$ ct cm$^{-2}$ s$^{-1}$ day$^{-1}$ ($\chi^2/\nu$ =
0.33/1).}
\label{lc}
\end{figure}

The three-point X-ray lightcurve from the 3 ACIS-I detections of s1-86
is shown in Figure~\ref{lc}.  Although there are only 3 data points,
the decay is not consistent with an exponential.  The best fitting
exponential decay is shown on the figure with a dashed line.  It has
$\chi^2/\nu=27.28/1$ and an $e$-folding time of 26.9 days.  On the
other hand, a linear decay, shown with the solid line, provides an
excellent fit to the data ($\chi^2/\nu=0.33/1$) with a slope of -9.3
$\times$10$^{-6}$ ct cm$^{-2}$ s$^{-1}$ day$^{-1}$ and a {\it total}
decay time of $\sim$60 days.

To look for short-term variability, we performed a Kolmogorov-Smirnov
(KS) test between the photon arrival times for s1-86 during its first
detection (ObsID 4678) and a uniform photon arrival distribution
(simulating a constant source) containing the same number of counts.
We considered the arrival times for the 758 counts contained in the
region of s1-86 as measured by {\it wavdetect}.  This test returned a
P-value of 0.969, which provides no indication of variability during
the observation.

Next, the three positional measurements for s1-86 and their errors
(given in Table~\ref{xpos}) were combined to determine the most
accurate coordinates for the X-ray source.  Table~\ref{xpos} includes
the 3 largest sources of positional errors in our analysis.  These
errors were all included in our determination of the X-ray position of
s1-86.  The quantity $\sigma_{cen}$ corresponds to the positional
error measured by {\it imcentroid}, which is limited by the large PSF
at the location of s1-86, 16$'$ off-axis.  The quantity $\sigma_{AL}$
corresponds to the positional error of the alignment of the ACIS X-ray
positions with the LGS coordinate system, measured by {\it ccmap}. The
quantity $\sigma_{S\rightarrow I}$ corresponds to the positional error
of sources on the ACIS-S chips when ACIS-I is the aimpoint, as was the
case for all 3 of our detections of s1-86; this error was estimated by
Markevitch.\footnote{http://asc.harvard.edu/cal/Hrma/optaxis/platescale/geom\_public.html}
He found an error of 1$''$ is typical for such positions.  This error
is independent of our alignment procedure as none of our alignment GCs
were located on the S-chips.  We included this error by adding it
directly to our totals from the other errors.

The weighted mean of the three positions and combined random errors
(excluding $\sigma_{S\rightarrow I}$) provided our best measurement of
the location of s1-86: R.A.=00:44:06.68 $\pm$0.74$''$, Dec.=41:12:20.0
$\pm$1.31$''$.  After adding the systematic ($\sigma_{S\rightarrow
I}$) error to both coordinates, the final (1$\sigma$) R.A. position
error is 1.74$''$, and the final (1$\sigma$) Dec. position error is
2.31$''$.  At the distance of M31 (780 kpc; \citealp{williams2003}),
these errors correspond 6.6 pc and 8.7 pc.

In addition, as a consistency check to our weighted mean position, we
co-added the 3 aligned ACIS observations and measured the {\it
wavdetect} position and {\it imcentroid} position errors of the
combined image of s1-86.  This position is R.A.=00:44:06.75
$\pm$0.61$''$, Dec.=41:12:20.2 $\pm$1.35$''$, in good agreement with
the weighted mean position calculated above.  After adding the
systematic ($\sigma_{S\rightarrow I}$) error to both coordinates, the
final (1$\sigma$) R.A. position error is 1.61$''$, and the final
(1$\sigma$) Dec. position error is 2.35$''$.

Finally, the three X-ray spectra were all well-fit by an absorbed
power-law model and reasonably fit (though not as well) by an absorbed
disk blackbody model.  All of the properties and $\chi^2$ values for
the fits are provided in Tables~\ref{spec.dat} and \ref{spec.dat2}. 

The power-law fits show no evidence for spectral variability during
the decay of the X-ray flux.  All of the power-law fits are consistent
with a power-law index of 3 and an absorption column of
2$\times$10$^{21}$ cm$^{-2}$.  The disk blackbody fits suggest that,
during the decay of the X-ray flux, there was a decrease in the inner
disk temperature from 0.6 keV to 0.4 keV.  These fits yield a constant
inner disk radius $\gap$50 km and a constant absorption column of
5$\times$10$^{20}$ cm$^{-2}$.  The temperatures and radii are similar
to those found for ACIS-I observations of the M31 X-ray transient
r2-67 ($T_{in}$ = 0.35$\pm$0.05 keV, $R_{in} cos^{1/2}i \gap 140$ km),
which had an optical counterpart detected with {\it HST}
\citep{williams2004}.

The best estimate of the absorption to s1-86 is the weighted mean of
the values measured from the first 2 spectral measurements, which is
(2.2$\pm$0.3)$\times$10$^{21}$ cm$^{-2}$ (power-law model) or
(5.2$\pm$0.9)$\times$10$^{20}$ cm$^{-2}$ (disk blackbody model).  The
maximum observed absorption-corrected luminosity is 0.9--2.4
$\times$10$^{38}$ erg s$^{-1}$ depending on the model fit, so the
luminosity alone does not discriminate between an accreting black hole
or neutron star.

\subsection{Optical}

We searched the ACS images, aligned to the same coordinate system as
the ACIS images, for sources within 4.5$''$ (2$\sigma$) of the X-ray
position.  DAOPHOT found 791 optical sources in the first ACS
observation down to $B$=27.8 (the 4$\sigma$ detection limit).  DAOPHOT
found 799 optical sources in second ACS observation down to $B$=27.8
within 4.5$''$ of the X-ray position.  With a distance modulus to M31
of 24.47 and a typical extinction $A_B\sim0.4$, this detection limit
corresponds to M$_B\sim2.9$.

In addition, we note that the brightest star within 4.5$''$ of the
X-ray position has $B=24.25\pm0.03$, fainter than a counterpart for a
high-mass Be transient.  Such transients have optical counterparts
with spectral type O8.5--B2 \citep{negueruela1998}, $-5.0<M_B<-2.6$,
or $19.8.0<B<22.2$ scaled to the distance and extinction of M31.

None of the optical sources down to $B$=26.8 within the 1$\sigma$
error ellipse of the X-ray source position in the first ACS image had
a significantly lower ($\Delta m\geq$1 mag) optical flux in the second
ACS observation.  Extending the search to the 2$\sigma$ error ellipse,
there is only 1 optical source with $B\leq$26.8 in the first ACS image
that had a significantly lower ($\Delta m\geq$1 mag) optical flux in
the second ACS observation.  This sole counterpart candidate is
located at R.A.=00:44:06.81 and Dec.= +41:12:24.0 (J2000).  This
position is shown on the ACS images of the region of interest in
Figure~\ref{hstims}.  The source flux decreased by 70\%, from
$m_B=25.75\pm0.05$ to $m_B=27.1\pm0.1$, between ACS observations.
Further details about this source in both ACS observations are
provided in Table~\ref{hst.dat}.

If this fading optical source is indeed the counterpart of s1-86, we
can use the simultaneous X-ray and optical luminosities to predict the
orbital period for the system using the relation of
\citet{vanparadijs1994}. Since this relation was seen in Galactic XRNe
events, there have been many new systems discovered.  Using the LMXB
catalog of \citet{liu2001}, we checked the consistency of this
relation with more recent X-ray novae, including GRO~J1655-40,
4U~1543-47, XTE~1550-564, XTE~J1118+480, and V4641~Sgr.  With the
exception of XTE~J1118+480, which was two magnitudes brighter in the
optical than the prediction from the relation, these more recent XRNe
all fit the \citet{vanparadijs1994} relation.  The higher than
expected optical luminosity of XTE~J1118+480 can be understood in the
context of the van Paradijs and McClintock model, which postulates
that the optical flux is due solely to X-ray heating.  XTE~J1118+480
has the lowest X-ray luminosity of all sources in the sample, and is
therefore the most likely to reveal optical emission above and beyond
that due to X-ray heating.

The \citet{vanparadijs1994} relation requires only simultaneous
measurements of the absolute $V$ magnitude and the X-ray luminosity to
predict the orbital period of an LMXB system.  These measurements can
be taken from any part of an outburst when the dominant contributor to
the optical flux is X-ray heating; the peak luminosity and outburst
amplitude are not required in the application of the relation.

We determined the L$_X$/L$_{opt}$ ratio at the time of the first $HST$
observation (03-Dec-2003), assuming the X-ray luminosities given by
the power-law fits in Table~\ref{spec.dat}.  We apply the X-ray
luminosity on 03-Dec-2003 as predicted by our linear decay fit to the
lightcurve (1.4$\times$10$^{38}$ erg s$^{-1}$).  To determine $M_V$,
we apply the $B$ magnitude of our candidate (25.75$\pm$0.05), and
correct for the absorption measured in the X-ray spectra
((2.2$\pm$0.3) $\times$10$^{21}$ cm$^{-2}$).  This absorption can be
converted to $A_B=1.6\pm0.2$ using the conversion of
\cite{predehl1995} and assuming a standard interstellar extinction
law.  These values give $M_B = -0.34\pm0.23$.  Assuming the mean
intrinsic B-V colors of LMXBs in the \cite{liu2001} Galactic LMXB
catalog (-0.09$\pm$0.14), $M_V = -0.25\pm0.27$ for the optical
counterpart candidate.

Applying these L$_X$ and $M_V$ values and errors to the relation of
\cite{vanparadijs1994}, including the errors on the parameters of the
relation, we obtain an orbital period of 1.0$^{+2.9}_{-0.6}$ day.
Applying Kepler's law, the inferred disk size is then
$\leq$0.7$^{+0.9}_{-0.4}\ \times 10^{12} \ M_{10}^{1/3}$~cm.  These
values are reasonable, but pushing the low end, for a large disk that
would produce a long, linear X-ray decay like the one observed for
s1-86 \citep{king1998}.

If we simply apply the X-ray luminosity and absorption given by the
power-law fit from the X-ray detection closest in time to the first
optical detection (26-November-2003; 1.8$\times$10$^{38}$ erg
s$^{-1}$; see Table~\ref{spec.dat}), rather than interpolating the
X-ray luminosity on December 3 as we have done above, the calculated
period is not significantly affected (0.9$^{+2.3}_{-0.7}$ day).
However, if we apply the X-ray luminosity and absorption column values
given by the disk blackbody fits in Table~\ref{spec.dat2}, the
calculated period is 0.3$^{+0.5}_{-0.1}$ day.  Applying Kepler's law,
the inferred disk size is then $\leq$3$^{+2}_{-1}\ \times 10^{11} \
M_{10}^{1/3}$~cm.

If our only plausible optical counterpart candidate is not the true
counterpart, then the true counterpart must have been fainter than
$B=26.8$.  Since our (4$\sigma$) detection limit was $B$=27.8, a
source fainter than $B=26.8$ in the first image would only provide a
lower-limit for the magnitude decrease, and the lower-limit would be
less than 1 mag, even if the source was unseen in the second ACS
observation.  Applying the same calculations as described above to
this conservative $B$ magnitude limit (if the optical candidate is not
the true counterpart) provides an upper-limit on the orbital period of
the system of $p<0.6$ day and an upper-limit on the size of the disk
of $a < 4.5\times10^{11}\ M_{10}^{1/3}$ cm.  Such values are below
those predicted for XRNe with long, linear X-ray decays
(\citealp{king1998}; see \S~4.1.2).

\section{Discussion}

\subsection{Arguments for an optical counterpart}\label{pro}

\subsubsection{Optical magnitudes}

\citet{chen1997} determined that the mean optical decay timescale for
Galactic XRNe is a factor of 2.2 longer than the X-ray decay
timescale.  This mean difference in decay timescales is in agreement
with theoretical models of accretion processes, which predict that the
optical decay timescales will be longer than X-ray decay times by a
factor of 2--4 \citep{king1998}.

For s1-86, the {\it total} X-ray decay time was $\sim$60 days. Since
the optical flux of the second detection (01-Mar-2004) of the
counterpart candidate was 30\% of the 03-Dec-2003 optical flux 88 days
earlier, we estimate the optical decay time for the candidate to be
$\sim$130 days.  Our estimate of the optical/X-ray decay time ratio is
therefore $\sim$2.2.  

The decay time ratios for the Galactic XRNe in the \citet{chen1997}
sample cover a wide range, as do more recent Galactic XRNe.  In some
cases, such as the May 1989 XRNe of 2023+338 (V404 Cyg), the optical
decay time has been observed to be shorter than the X-ray decay time
(\citealp{kitamoto1989,wagner1991,chen1997}).  However, we note that
the optical decay of the source detailed in Table~\ref{hst.dat} at
R.A.=00:44:06.81, Dec.=41:12:24.0 (J2000) lies near the mean of
Galactic events and is consistent with theoretical models.  This
consistency argues in favor of the candidate being the true optical
counterpart.

\subsubsection{Disk size and orbital period}

As discussed in \citet{king1998}, one may use the decay time of the
XRN to predict the size of the accretion disk of the system.  This
method was applied to the X-ray transient source r2-67 by
\citet{williams2004} and gave an answer consistent with the period
prediction from the optical and X-ray luminosities.

We calculated the size of the disk expected for a 10$M_{\odot}$ black
hole with a linear decay to approximately zero X-ray flux 60 days
after a peak X-ray luminosity of 2.4$\times$10$^{38}$ erg s$^{-1}$.
Similar to the calculation for r2-67 in \citet{williams2004}, the
disk size prediction for s1-86 is
$$
2\times10^{12}{M_{10}^{1/3}\over \alpha^{2/3}T_4^{2/3}} \ \ cm,
$$ 
where $\alpha$ is the angular momentum transport efficiency, $M_{10}$
is the black hole mass in 10 $M_{\odot}$ units, and $T_4$ is the local
disk temperature in units of $10^4$ K.  We note that if the true peak
X-ray luminosity was higher, the disk size predicted by the model
would increase.  If we apply the lowest possible peak X-ray
luminosity, that given by the disk blackbody fit to the 09-Nov-2003
data (9$\times$10$^{37}$ erg s$^{-1}$), the disk size prediction is
$$
10^{12}{M_{10}^{1/3}\over \alpha^{2/3}T_4^{2/3}} \ \ cm.
$$ 

According to Kepler's laws, the size of the orbit of the secondary is
$a = 6.3\times10^{11} P_{1}^{2/3} M_{10}^{1/3}$ cm, where $P_{1}$ is
the orbital period in days.  Because the disk cannot be larger than
the orbit of the secondary ($a$), these calculations suggest that the
orbital period of this system is $>$1 day.  This estimate of the
orbital period is consistent with the prediction from the
\citet{vanparadijs1994} relation when the optical luminosity of the
counterpart candidate is applied (see \S~3.2), providing some
additional confidence that the candidate is the true optical
counterpart.

\subsection{Arguments against an optical counterpart}\label{anti}

\subsubsection{Positional (dis)agreement}

On the other hand, the precision of the X-ray position for s1-86 is
not optimal.  As shown in Table~\ref{xpos} and discussed in
\S~\ref{results}, the X-ray position errors include 50 square
arcseconds of area within the 2$\sigma$ error ellipse.

The optical candidate is 0.8$\sigma$ away from the R.A. of the X-ray
position and 1.7$\sigma$ away from the Declination of X-ray position.
Assuming Gaussian statistics, the optical source has $\sim$9\%
probability of occupying the same location as the X-ray source;
however, we do not rule out the possibility that this is the
counterpart because the {\it Chandra} PSF is known to be complicated
far off-axis.  In addition, as shown in Figure~\ref{xims}, the
location of the optical counterpart candidate appears qualitatively
reasonable when placed on the X-ray images of s1-86.

\subsubsection{Alternate optical source possibilities}

If the optical candidate is not the counterpart of the X-ray
transient, the candidate must be a different type of optical source
that varies by 1.3 mag in brightness.  Such sources are not extremely
rare.  We searched the overlapping portions of our 2 ACS observations
for other sources that became fainter by more than 1 mag between the
first and second observations.  We found 75 such sources in the 9.2752
square arcminutes of overlapping area.  Since our 2$\sigma$ error
ellipse covers 0.0140 square arcminutes, there is an 11\% random
probability of one of these sources falling in this region, similar to
the probability that the candidate occupies the same position as the
X-ray source.

There are several types of variable sources that could show the same
behavior as the counterpart candidate, including eclipsing binaries,
high-amplitude variable stars, and dwarf nova systems.  For example, a
common type of variable with properties similar to those of the
counterpart candidate is RR Lyrae stars.  M31 is known to have $\gap$5
RR Lyrae stars per square arcminute \citep{brown2004}.  At $B\sim26$,
assuming a distance modulus of 24.47 (780 kpc) and foreground
extinction ($A_B=0.4$), the counterpart candidate has $M_B\sim1$,
typical of RR$_{ab}$ Lyrae stars, which can vary by 0.5$\lap\Delta
m\lap$1.5 mag \citep{textbook}.  If the candidate is such a star, its
period and phase must have allowed us to observe it near maximum
during our first {\it HST} observation and near minimum during our
second {\it HST} observation.  There were 88.8 days between our two
{\it HST} observations.  If this optical source is an RR$_{ab}$ star,
its period is likely $\sim$0.55 days \citep{textbook}.  Therefore, the
star would have undergone 161.5 cycles between our observations,
allowing us to observe it in opposite phases.  It is clearly possible
that the counterpart candidate is an unassociated variable star.

\section{Conclusions}

We have reported the detection of a new, bright transient X-ray source
in M31, which we call s1-86 (CXOM31 J004406.7+411220).  After aligning
the {\it Chandra} ACIS-I images to the LGS coordinate system, this
source is located at R.A.=00:44:06.68, Dec.=41:12:20.0 with final
R.A. and Dec. errors of 1.74$''$ and 2.31$''$, respectively.  The
X-ray spectral properties of this source as measured from 3 {\it
Chandra} ACIS-I detections are typical of known Galactic XRNe.
Power-law fits indicate a spectral index of 3 and an absorption column
of 2.2$\times$10$^{21}$ cm$^{-2}$.  Disk blackbody fits indicate an
inner disk temperature decrease from 0.6 keV to 0.4 keV over 1 month
and an absorption column of 5$\times$10$^{20}$ cm$^{-2}$.  The 0.3--7
keV luminosity was $\sim$2.4$\times$10$^{38}$ erg s$^{-1}$
(power-law), or $\sim$0.9$\times$10$^{38}$ erg s$^{-1}$ (disk
blackbody), depending on the spectral model employed.  The X-ray
lightcurve exhibits a linear decay, suggesting a large accretion disk.

In two {\it HST} ACS observations, we detected only one optical source
within 4.5$''$ of the X-ray position of the XRN that faded in concert
with the X-ray source.  After aligning the ACS images to the LGS
coordinate system, this optical counterpart candidate was located at
R.A.=00:44:06.81 and Dec.= +41:12:24.0 (J2000).  While the optical
brightness and decay timescale suggest that this optical source is the
counterpart, the errors in the X-ray source position allow the
possibility that the counterpart candidate is a nearby RR$_{ab}$ Lyrae
star which we observed near maximum and minimum brightness.  If so, no
optical counterpart was seen down to $B=26.8$.

If the candidate is the optical counterpart and the X-ray luminosity
indicated by the power-law spectral fit is correct, then the
X-ray/optical luminosity ratio yields predictions for the orbital
period of the X-ray binary and the size of the accretion disk.  The
period prediction is 1.0$^{+2.9}_{-0.6}$ day, and the disk size
prediction is $\leq$0.7$^{+0.9}_{-0.4}\times 10^{12} M_{10}^{1/3}$~cm.
These values are on the low side of model predictions for XRNe with
long, linear X-ray decays, and the lower X-ray luminosity given by the
disk blackbody spectral fit yields even lower values.

If the optical counterpart was not detected, it was at least 1 mag
fainter than the candidate counterpart.  This upper-limit to the
optical luminosity provides lower-limits on the X-ray/optical
luminosity ratios given by both X-ray spectral models.  In both cases,
these lower-limits yield upper-limits on the orbital period and the
disk size that are below model predictions for XRNe with long, linear
X-ray decays.

Support for this work was provided by NASA through grant number
GO-9087 from the Space Telescope Science Institute and through grant
number GO-3103X from the {\it Chandra} X-Ray Center.  MRG acknowledges
support from NASA LTSA grant NAG5-10889.


\clearpage

\begin{deluxetable}{ccccccccccc}
\tablecaption{{\it Chandra} ACIS-I observations}
\tableheadfrac{0.01}
\tablehead{
\colhead{{ObsID}} &
\colhead{{Date}} &
\colhead{{R.A. (J2000)}} &
\colhead{{Dec. (J2000)}} &
\colhead{{Roll (deg.)}} &
\colhead{{Exp. (ks)}} 
}
\startdata
4678 & 09-Nov-2003 & 00 42 44.4 & 41 16 08.3 & 239.53 & 3.9\\
4679 & 26-Nov-2003 & 00 42 44.4 & 41 16 08.3 & 261.38 & 3.8\\
4680 & 27-Dec-2003 & 00 42 44.4 & 41 16 08.3 & 285.12 & 4.2\\
\enddata
\label{xobs}
\end{deluxetable}

\begin{deluxetable}{ccccccccccc}
\tablewidth{7in}
\tablecaption{Position Measurements and Errors of XRN s1-86}
\tableheadfrac{0.01}
\tablehead{
\colhead{{ID}} &
\colhead{{R.A. (J2000)}} &
\colhead{{$\sigma_{cen}$}} &
\colhead{{$\sigma_{AL}$}} &
\colhead{{$\sigma_{S\rightarrow I}$}} &
\colhead{{$\sigma_{tot}$}} &
\colhead{{Dec. (J2000)}} &
\colhead{{$\sigma_{cen}$}} &
\colhead{{$\sigma_{AL}$}} &
\colhead{{$\sigma_{S\rightarrow I}$}} &
\colhead{{$\sigma_{tot}$}} 
}
\startdata
4678 & 00 44 06.64 & 0.96$''$ & 0.10$''$ & 1$''$ & 1.96$''$ & 41 12 20.0 & 1.68$''$ & 0.18$''$ & 1$''$ & 2.69$''$\\
4679 & 00 44 06.73 & 1.23$''$ & 0.09$''$ & 1$''$ & 2.23$''$ & 41 12 20.1 & 2.40$''$ & 0.19$''$ & 1$''$ & 3.41$''$\\
4680 & 00 44 06.85 & 3.79$''$ & 0.12$''$ & 1$''$ & 4.80$''$ & 41 12 19.9 & 3.96$''$ & 0.24$''$ & 1$''$ & 4.96$''$\\
\hline
Mean & 00 44 06.68 & \nodata & \nodata & \nodata & 1.74$''$ & 41 12 20.0 & \nodata & \nodata & \nodata & 2.31$''$\\
\enddata
\label{xpos}
\end{deluxetable}


\begin{deluxetable}{cccccccc}
\tablewidth{6.5in}
\tablecaption{{\it HST} ACS F435W photometry for the candidate optical counterpart of s1-86}
\tableheadfrac{0.01}
\tablehead{
\colhead{{Date}} & 
\colhead{{R.A.(J2000)}} & 
\colhead{{Dec.(J2000)}} & 
\colhead{{($''$)\tablenotemark{a}}} &
\colhead{{Rate (ct s$^{-1}$)}} &
\colhead{{STmag\tablenotemark{b}}} &
\colhead{{VEGAmag\tablenotemark{c}}} 
}
\tablenotetext{a}{Distance between this source position and the X-ray source position.}
\tablenotetext{b}{The ST magnitude of the source is obtained from a system with a flat reference spectrum (see http://www.stsci.edu/hst/acs/documents/handbooks/DataHandbookv2 for more details).}
\tablenotetext{c}{The VEGA magnitude of the source is obtained from a system where Vega has a magnitude of 0 at all wavelengths.  This magnitude is the best approximation of the Johnson $B$ magnitude of the source.}
\startdata
03-Dec-2003 & 00:44:06.81 & +41:12:24.0 & 4.26 & 1.00$\pm$0.05 & 25.14$\pm$0.05 & 25.75$\pm$0.05\\
01-Mar-2004 & 00:44:06.81 & +41:12:24.0 & 4.26 & 0.30$\pm$0.03 & 26.47$\pm$0.11 & 27.08$\pm$0.11\\
\enddata
\label{hst.dat}
\end{deluxetable}

\clearpage

\begin{deluxetable}{ccccccccccc}
\tablewidth{6in} 
\tablecaption{Hardness Ratios and Power-law Spectral Fits to ACIS-I Detections of XRN s1-86} 
\tableheadfrac{0.01} 
\tablehead{ 
\colhead{{\tiny{Date}}} &
\colhead{{\tiny{ObsID}}} & 
\colhead{{\tiny{Cts\tablenotemark{a}}}} &
\colhead{{\tiny{Flux\tablenotemark{b}}}} & 
\colhead{{\tiny{Slope\tablenotemark{c}}}} & 
\colhead{{\tiny{${\rm N_H}$}}} & 
\colhead{{\tiny{$\chi^2/\nu$}}} &
\colhead{{\tiny{Q\tablenotemark{d}}}} & 
\colhead{{\tiny{HR-1\tablenotemark{e}}}} &
\colhead{{\tiny{HR-2\tablenotemark{f}}}} & 
\colhead{{\tiny{$\rm L_X$\tablenotemark{g}}}}
} 
\tablenotetext{a}{The background-subtracted number of counts in the detection.}
\tablenotetext{b}{The exposure corrected 0.3--10 keV flux in units of 10$^{-4}$ ct cm$^{-2}$ s$^{-1}$.}
\tablenotetext{c}{Slope of the best-fitting absorbed power law model.}
\tablenotetext{d}{The probability that this fit is representative of
the true spectrum, determined from $\chi^2/dof$.}
\tablenotetext{e}{Hardness ratio calculated by taking the ratio of
M-S/M+S, where S is the number of counts from 0.3--1 keV and M is the
number of counts from 1--2 keV.}  
\tablenotetext{f}{Hardness ratio
calculated by taking the ratio of H-S/H+S, where S is the number of
counts from 0.3--1 keV and H is the number of counts from 2--7 keV.}
\tablenotetext{g}{The absorption-corrected luminosity of the source in
units of 10$^{38}$ \ergs (0.3--7 keV).}  
\tablenotetext{h}{The absorption was fixed to fit the spectrum from 27-Dec-2003, as $N_H$ was unconstrained if left as a free parameter.} 
\startdata
\tiny{09-Nov-2003} & \tiny{4678} & \tiny{666} & \tiny{5.1$\pm$0.2} & \tiny{2.9$\pm$0.2} & \tiny{2.25$\pm$0.35} & \tiny{39.33/30} & \tiny{0.12} & \tiny{0.11$\pm$0.04} & \tiny{-0.58$\pm$0.07} & \tiny{2.4}\\
\tiny{26-Nov-2003} & \tiny{4679} & \tiny{465} & \tiny{3.6$\pm$0.2} & \tiny{3.0$\pm$0.3} & \tiny{2.13$\pm$0.47} & \tiny{21.40/19} & \tiny{0.31} & \tiny{0.02$\pm$0.05} & \tiny{-0.69$\pm$0.08} & \tiny{1.8}\\ 
\tiny{27-Dec-2003} & \tiny{4680} & \tiny{79} & \tiny{0.7$\pm$0.1} & \tiny{3.4$\pm$0.4} & \tiny{2.0\tablenotemark{h}} & \tiny{3.47/8} & \tiny{0.90} & \tiny{0.10$\pm$0.11} & \tiny{-1.14$\pm$0.39} & \tiny{0.5}\\ 
\enddata
\label{spec.dat}
\end{deluxetable}


\oddsidemargin 0.0cm

\clearpage

\begin{deluxetable}{ccccccccccc}
\tablewidth{6.5in} \tablecaption{Disk Blackbody Spectral Fits to ACIS-I Detections of XRN s1-86} 
\tableheadfrac{0.01} 
\tablehead{ 
\colhead{{Date}} &
\colhead{{OBSID}} & 
\colhead{{T$_{in}$\tablenotemark{a}}} & 
\colhead{{R$_{in}cos^{1/2}i$ \tablenotemark{b}}} &
\colhead{{${\rm N_H}$}} & 
\colhead{{$\chi^2/\nu$}} &
\colhead{{Q\tablenotemark{c}}} & 
\colhead{{$\rm L_X$\tablenotemark{d}}} } 
\tablenotetext{a}{The temperature of the inner disk in keV.}
\tablenotetext{b}{The radius of the inner disk in km, assuming the distance to M31 is 780 kpc and the inclination ($i$) of the binary is 0 degrees.}
\tablenotetext{c}{The probability that this fit is representative of
the true spectrum, determined from $\chi^2/dof$.}
\tablenotetext{d}{The absorption-corrected luminosity of the source in
units of 10$^{38}$ \ergs (0.3--7 keV).}  
\tablenotetext{e}{The absorption was fixed to fit the spectrum from 27-Dec-2003, as $N_H$ was unconstrained if left as a free parameter.} 
\startdata
09-Nov-2003 & 4678 & 0.60$\pm$0.04 & 56$\pm$9 & 0.5$\pm$0.1 & 39.37/30 & 0.12 & 0.9\\ 
26-Nov-2003 & 4679 & 0.49$\pm$0.05 & 74$^{+15}_{-19}$ & 0.6$\pm$0.2 & 30.37/19 & 0.05 & 0.6\\ 
27-Dec-2003 & 4680 & 0.37$\pm$0.07 & 65$^{+24}_{-40}$ & 0.5\tablenotemark{e} & 4.51/8 & 0.81 & 0.2\\ 
\enddata
\label{spec.dat2}
\end{deluxetable}

\clearpage

\end{document}